\documentclass[aps,pra,superscriptaddress,12pt,tightenlines,nofootinbib]{revtex4}
\usepackage{amsmath,amsthm,graphicx,amssymb,verbatim,listings}
\usepackage{wasysym}
\usepackage[T1]{fontenc}     % needed for the guillemets
\usepackage{lmodern}         % needed to fix suboptimal font rendering
\usepackage[ pdftex, plainpages = false, pdfpagelabels,
                 pdfpagelayout = useoutlines,
                 bookmarks,
                 bookmarksopen = true,
                 bookmarksnumbered = true,
                 breaklinks = true,
                 linktocpage=all,
                 pagebackref=false,
                 colorlinks = true,
                 linkcolor = BrickRed,
                 urlcolor  = blue,
                 citecolor = BrickRed,
                 anchorcolor = green,
                 hyperindex = true,
                 hyperfigures
                 ]{hyperref}
\usepackage[usenames, dvipsnames]{xcolor}
\usepackage{xifthen} % provides \isempty test

\newcommand{\tr}{\hbox{tr}}

\newcommand{\ket}[1]{{\ensuremath{\left| #1 \right\rangle}}}
\newcommand{\bra}[1]{{\ensuremath{\left\langle #1 \right|}}}
\newcommand{\braket}[2]{{\ensuremath{\left\langle #1 \middle| #2
      \right\rangle}}}
\newcommand{\ketbra}[2]{{\ensuremath{\left| #1 \middle\rangle\!\middle\langle #2
      \right|}}}

\newcommand{\arxiv}[2][]{\ifthenelse{\isempty{#1}}{\href{http://arxiv.org/abs/#2}{{\tt arXiv:\allowbreak{}#2}}} {\href{http://arxiv.org/abs/#2}{{\tt arXiv:\allowbreak{}#2 [#1]}}}}
\newcommand{\pirsa}[1]{\href{http://pirsa.org/#1/}{{\tt PIRSA:\allowbreak{}#1}}}
\newcommand{\booktitle}{\textsl}
\newcommand{\hrefdoi}[2]{\href{https://dx.doi.org/#1}{#2}}

\begin{document}

\title{Invariant Off-Diagonality:\ SICs as Equicoherent Quantum States}

\author{Blake C.\ Stacey}
\affiliation{\href{http://www.physics.umb.edu/Research/QBism/}{QBism Research Group}, Physics Department \\ University of Massachusetts Boston}

\date{\today}

\begin{abstract}
  Coherence, treated as a resource in quantum information theory, is a
  basis-dependent quantity. Looking for states that have constant
  coherence under canonical changes of basis yields highly symmetric
  structures in state space. For the case of a qubit, we find an easy
  construction of qubit SICs (Symmetric Informationally Complete
  POVMs). SICs in dimension 3 and 8 are also shown to be equicoherent.
\end{abstract}

\maketitle

In practice it may be helpful to think of a given set of quantum
states as representing laboratory procedures that are easy to do.  If
the vectors comprising one orthonormal basis represent preparations
that are convenient or inexpensive, then it is reasonable to say that
density matrices that are strongly off-diagonal in that basis
correspond to preparations that are more costly.  We might then ask,
for example, what tasks become practical if we can carry out one
costly preparation and an arbitrarily large number of cheap
transformations~\cite{Streltsov:2017}.  Having put ourselves in the
mindset of viewing off-diagonality as a \emph{resource,} we encounter
a natural generalization: What if we have a choice of inexpensive
bases?  For example, we might be dealing with transmission errors that
stochastically flip which basis is cheap~\cite{Bennett:1997}.  Are
there preparation procedures that are \emph{equally costly} with
respect to any one of a canonical discrete set of bases?

To make the question concrete, take the case of a single qubit. A
quantum state that can be ascribed to a qubit-sized system is a
$2\times2$ positive semidefinite matrix of unit trace, which we can
neatly express as a sum over Pauli operators:
\begin{equation}
  \rho = \frac{1}{2} (I + x\sigma_x + y\sigma_y + z\sigma_z)
  = \frac{1}{2} \left(\begin{matrix}
    1 + z & x - iy \\
    x + iy & 1 - z
    \end{matrix}\right).
\end{equation}
Here, $(x,y,z)$ are the coordinates of the state $\rho$ in the Bloch
sphere representation. A handy measure of how off-diagonal the state
$\rho$ is in the eigenbasis of the Pauli operator $\sigma_z$ is the
sum of the squared magnitudes of the off-diagonal entries, which is
\begin{equation}
  \frac{1}{4}|x - iy|^2 + \frac{1}{4}|x + iy|^2
  = \frac{1}{2}(x^2 + y^2).
\end{equation}
What states are equally off-diagonal by this measure in the eigenbases
of $\sigma_x$, $\sigma_y$ and $\sigma_z$?  The coordinates of such a
state must satisfy
\begin{equation}
  x^2 + y^2 = x^2 + z^2 = y^2 + z^2.
\end{equation}
Let us confine our attention to pure states, which lie on the surface
of the Bloch ball:
\begin{equation}
  x^2 + y^2 + z^2 = 1.
\end{equation}
Combining these constraints, we find that
%\begin{equation}
%  z^2 = 1 - (x^2 + y^2) = 1 - (x^2 - z^2) = y^2 = 1 - (y^2 - z^2) = x^2.
%\end{equation}
%Therefore,
\begin{equation}
  x^2 = y^2 = z^2 = \frac{1}{3},
\end{equation}
meaning that the states we seek are the vertices of a cube inscribed
in the Bloch sphere:
\begin{equation}
  x,y,z \in \left\{ \pm \frac{1}{\sqrt{3}} \right\}.
\end{equation}
This set of eight states naturally breaks down into two sets of four,
which are orbits under the action of the Pauli group.  Each set of
four states forms a tetrahedron inscribed in the Bloch sphere, with
the vertices of one tetrahedron antipodal to those of the other.  One
tetrahedron comprises the sign choices of even parity, and the other
the sign choices of odd parity.  We conclude that the set of qubit
pure states that are equicoherent in all three Pauli bases are the
\emph{qubit SIC states.}

A SIC, we recall, is a set of $d^2$ unit vectors $\{\ket{\pi_j}\}$ in
$\mathbb{C}^d$ that enjoy the symmetry property
\begin{equation}
  |\braket{\pi_j}{\pi_k}|^2 = \frac{d\delta_{jk}+1}{d+1}.
\end{equation}
The acronym stands for \emph{Symmetric Informationally Complete,}
referring to the measurement that is formed by scaling each of the
projectors $\ketbra{\pi_j}{\pi_j}$ by $1/d$. For the history and
significance of these entities, we refer to the
literature~\cite{Zauner:1999, Renes:2004, Scott:2010, Bengtsson:2017,
  Appleby:2017, Fuchs:2017, Waldron:2018, Kopp:2018, DeBrota:2018a,
  DeBrota:2018b}.

Note that if we had used the ``$l_1$-norm of
coherence''~\cite{Zhu:2018} instead, we would have found the
constraint
\begin{equation}
  \sqrt{x^2 + y^2} = \sqrt{x^2 + z^2} = \sqrt{y^2 + z^2},
\end{equation}
to ultimately the same effect.

In any finite dimension, pure quantum states satisfy
\begin{equation}
  \tr \rho = \tr \rho^2 = \tr \rho^3 = 1.
\end{equation}
Since the trace of $\rho^2 = \rho^\dag \rho$ is the sum of the squared
magnitues of all the elements of~$\rho$, we can relate the above
measure of off-diagonality to the diagonal entries:
\begin{equation}
  \sum_{i\neq j} |\rho_{ij}|^2 = 1 - \sum_i |\bra{i}\rho\ket{i}|^2.
  \label{eq:nice-sum}
\end{equation}
The numbers $\{\bra{i}\rho\ket{i}\}$ are, of course, the probabilities
ascribed to the outcomes of a measurement in the orthonormal basis
$\{\ket{i}\}$.

Suppose that the dimension $d$ is a power of a prime, so that a
complete set of $d+1$ Mutually Unbiased Bases (MUB) is known to exist.
Let $\ket{m,j}$ be the states comprising these MUB, with $m$ labeling
the basis and $j$ the vector within that basis. Given a state $\rho$,
we apply the Born rule to compute the probabilities
\begin{equation}
  p_{m,j} = \bra{m,j} \rho \ket{m,j}.
\end{equation}
From the fact that a complete set of MUB states forms a 2-design, it
follows that
\begin{equation}
  \sum_{m,j} p_{m,j}^2 = 2.
\end{equation}
A \emph{minimum uncertainty state}~\cite{Appleby:2014} distributes
this sum equally over all the $d+1$ bases:
\begin{equation}
  \sum_j p_{m,j}^2 = \frac{2}{d+1}\ \forall\ m.
\end{equation}
We see that any minimum uncertainty state will be equally off-diagonal
in all $d+1$ of the MUB.  This generalizes the result we found above,
since the states of SICs that are generated as orbits of the
Weyl--Heisenberg group are minimum uncertainty
states~\cite{Appleby:2014}.

The term ``coherence'' is rather drastically overloaded, having
different meanings in multiple fields, with SICs being important for
many of them. They are significant for ``coherence'' in the Dutch-book
and frame-theoretic senses of the word~\cite{Fuchs:2013, Bodmann:2017,
  Fuchs:2019}, and now we see that they are so in the ``quantum
coherence as a resource'' sense as well.

The literature is replete with alternative ways of quantifying the
off-diagonality of quantum states.  (My impression is that some
definitions lead to measures that might have more physical relevance,
while others are easier to calculate, and sometimes the practical
thing to do is try and use the latter to get a bound on the former.)
Another such measure has an information-theoretic flavor and is known
as the \emph{relative entropy of coherence.}  First, we define a
``dephasing'' operator that Procrusteanizes a density operator into a
basis:
\begin{equation}
  \Delta(\rho) = \sum_i (\bra{i} \rho \ket{i})\ketbra{i}{i}.
\end{equation}
The relative entropy of coherence for a state $\rho$ is the change in
von Neumann entropy between its original and dephased forms:
\begin{equation}
  C_r(\rho) = S(\Delta(\rho)) - S(\rho).
  \label{eq:rel-ent}
\end{equation}
We focus our attention on pure states, for which the latter term
vanishes and the relative entropy of coherence reduces to a simple
Shannon functional of the probabilities $\{ \bra{i} \rho \ket{i} \}$.

A \emph{MUB-balanced} state is one for which the Born-rule
probabilities for the measurements corresponding to different bases
are the same up to permutations~\cite{Wootters:2007, Appleby:2014b}.
For any bases $m$ and $m'$,
\begin{equation}
  p_{m,j} = p_{m',j'}
\end{equation}
for some index $j'$.  The Shannon functional is indifferent to
permutations of probability vectors, and so MUB-balanced states are
equicoherent across the MUB with respect to the relative entropy of
coherence.  Wootters and Sussman demonstrated that MUB-balanced states
are minimum uncertainty states~\cite{Wootters:2007}, implying that
they are also equicoherent with respect to the
sum-of-squared-magnitudes definition of coherence. We can see this
from Eq.~(\ref{eq:nice-sum}), since the sum over squared probabilities
does not depend upon their ordering.

In particular, the nine states of the \emph{Hesse SIC} in $d = 3$ are
all MUB-balanced. For each state in the Hesse SIC and each basis $m$,
$p_{m,j}$ is some permutation of the tuple $(0, \frac{1}{2},
\frac{1}{2})$. The combinatorics and finite geometry that make this
pattern possible also yield a Kochen--Specker proof for
qutrits~\cite{Bengtsson:2012, Stacey:2016c, Stacey:2018b} and are
relevant for identifying the Hesse SIC states as \emph{maximally magic
  resources} for quantum computation~\cite{Veitch:2014}.

The \emph{Hoggar-type SICs} in dimension 8 are sets of 64 states
constructed as orbits of the three-qubit Pauli
group~\cite{Hoggar:1998, Szymusiak:2016, Stacey:2016}. A convenient
starting point is the vector
\begin{equation}
  \ket{\pi_0} \propto (-1+2i,1,1,1,1,1,1,1)^{\rm T}.
\end{equation}
Taking the orbit of this state under the three-qubit Pauli group
yields a set of 64 equiangular complex lines.  Other choices of
initial vector are possible, but all the SICs found as orbits of the
three-qubit Pauli group are equivalent up to unitary or antiunitary
conjugations. We designate all SICs constructed in this way as
Hoggar-type SICs, of which the one generated from $\ket{\pi_0}$ is the
prototype on which we will focus our attention.

Written in terms of rank-1 projectors, Hoggar SIC states satisfy
\begin{equation}
  \tr \Pi_j \Pi_k = \frac{1}{9}
\end{equation}
whenever $j \neq k$.  Like the qubit and Hesse SIC states, these have
been identified as resources for quantum
computation~\cite{Campbell:2017}. Without explicit calculation, we can
already see that these states will display degeneracies among their
coherences. The relative entropy of coherence for the state $\Pi_0$ is
the von Neumann entropy of the ``dephased'' state
\begin{equation}
  \Delta(\Pi_0) = \sum_{i=0}^8 (\bra{i} \Pi_0 \ket{i})\ketbra{i}{i}.
\end{equation}
But if we use a canonical set of MUB, each of the vectors
$\{\ket{i}\}$ is defined as a simultaneous eigenstate of multiple
three-qubit Pauli operators~\cite{Lawrence:2002,
  Romero:2005}. Therefore, if $U$ is a three-qubit Pauli unitary of
which $\{\ket{i}\}$ are eigenvectors,
\begin{equation}
  \Delta(\Pi_0) = \sum_{i=0}^8 \left(\bra{i} U^\dag \Pi_0 U
  \ket{i}\right)\ketbra{i}{i}
   = \Delta(U^\dag \Pi_0 U).
\end{equation}
Because $U$ is an element in the same group whose action generates the
SIC,
\begin{equation}
  \Delta(\Pi_0) = \Delta(\Pi_j)
\end{equation}
for some value of $j$.  So, for each choice from the $d + 1 = 9$ MUB,
seven other SIC states will ``dephase'' to the same mixed state as
$\Pi_0$ does.

The group covariance of the SIC set implies that for any $\Pi_j$,
\begin{equation}
  \Delta(\Pi_j) = \Delta(D_j \Pi_0 D_j^\dag)
\end{equation}
for some three-qubit Pauli operator $D_j$.  The set of all unitaries
that map a Hoggar-type SIC to itself is a subset of the three-qubit
Clifford group. Moreover, the three-qubit Clifford group maps the MUB
states to each other. The von Neumann entropy of the ``dephased''
state $S(\Delta(\rho))$ depends only upon the values $\{ \bra{i} \rho
\ket{i}\}$.  If we fix $\ket{i'} = D_j^\dag \ket{i}$, then
\begin{equation}
  \bra{i} \Pi_0 \ket{i} = \bra{i'} \Pi_j \ket{i'}.
\end{equation}
The relative entropy of coherence for $\Pi_0$ with respect to the
basis $\{\ket{i}\}$ is thus equal to that for $\Pi_j$ with respect to
the basis $\{\ket{i'}\}$.

We might plausibly guess that any state in a Hoggar-type SIC will turn
out to be equicoherent across all nine MUB as well, thanks to the
large size of its stabilizer group~\cite{Zhu:2012, Zhu:2015,
  Stacey:2017}. That is, there are 6,048 different Clifford unitaries
which map the set $\{\Pi_j\}$ to itself and satisfy $U\Pi_0U^\dag =
\Pi_0$.  Moreover, the symmetry group of a
Hoggar-type SIC is \emph{doubly transitive,} able to map any pair of
elements into any other.  Let $U$ be a Clifford unitary in the
stabilizer of $\Pi_0$, so that $U\Pi_0U^\dag = \Pi_0$ and $U \Pi_j
U^\dag = \Pi_k$.  Because $\Pi_j = D_j \Pi_0 D_j^\dag$ for some
three-qubit Pauli operator $D_j$, then
\begin{equation}
  U D_j \Pi_0 D_j^\dag U^\dag = D_k \Pi_0 D_k^\dag.
\end{equation}
But we can conjugate our state $\Pi_0$ by the stabilizer unitary $U$
and regroup:
\begin{equation}
  (U D_j U^\dag) \Pi_0 (U D_j^\dag U^\dag) = D_k \Pi_0 D_k^\dag.
\end{equation}
The only way it seems that this can work out is if $D_k$, which is
both unitary and Hermitian, is the same operator as that gotten by
conjugating $D_j$ with $U$.  From the double transitivity of the
Hoggar symmetry group, it follows that there must be a unitary in the
stabilizer of $\Pi_0$ that can turn any $\Pi_j$ into any desired
$\Pi_k$.  This in turn appears to require that the stabilizer of
$\Pi_0$ is transitive on the Pauli operators $\{D_j\}$.

The states of the Hoggar SIC are not MUB-balanced, but they are
minimum-uncertainty. By directly checking the overlaps with the
Wootters--Fields MUB states~\cite{Wootters:1989}, we find that the
probability distribution $\vec{p}_{m}$ is, for each basis, a
permutation either of the vector
\begin{equation}
  \left(\frac{5}{12}, \frac{1}{12}, \frac{1}{12}, \frac{1}{12},
  \frac{1}{12}, \frac{1}{12}, \frac{1}{12}, \frac{1}{12}\right)
\end{equation}
or of the vector
\begin{equation}
  \left(\frac{1}{3}, \frac{1}{6}, \frac{1}{6}, \frac{1}{6},
  \frac{1}{6}, 0, 0, 0\right).
\end{equation}
These two vectors have the same 2-norm, and so
\begin{equation}
  \sum_j p_{m,j}^2 = \frac{2}{9}
\end{equation}
for all 9 choices of $m$. In fact, for each of the 64 Hoggar states, 2
of the bases yield the first vector, and the other 7 bases yield the
second. This establishes equicoherence with respect to the definition
(\ref{eq:nice-sum}). It also establishes equicoherence with respect to
an information-theoretic measure like Eq.~(\ref{eq:rel-ent}), if the
entropy functional is the R\'enyi 2-entropy rather than the Shannon
formula~\cite{Zhu:2017}.

This example presents an intriguing generalization of the MUB-balanced
state concept: ``almost MUB-balanced'' quantum states, where there are
(up to permutations) two distinct probability vectors, both representing
``equal uncertainty''.

It also follows from the group covariance of a Hoggar-type SIC that
\begin{equation}
  \bra{\pi_j} D_k \ket{\pi_j} = \pm \frac{1}{3},
\end{equation}
where $D_k$ is any three-qubit Pauli operator.  Using these operators as
a Hermitian basis, we can write any $\Pi_j$ as a linear combination of
them, and the magnitudes of the coefficients in the expansion will be
the same for all $j$.  A generalization of the equicoherence property
from which we derived the qubit SIC states follows naturally.

From one perspective, coherent superpositions are not the deepest of
the quantum mechanical mysteries. It is possible to construct them in
theories that have underlying local hidden variables and that offer no
hope of computational speed-up. The idea in old books that
interference effects are quintessentially nonclassical is, in a modern
analysis, a failure of imagination~\cite{Spekkens:2007, Spekkens:2014,
  Spekkens:2016}. Useful as coherence may be for some protocols, it
does not appear to be the most potent resource within the scope of
quantum theory. Equicoherence, on the other hand, takes us out of that
intermediate, semiclassical regime.

\bigskip

I thank Marcus Appleby for discussions and John B.\ DeBrota for
verifying the situation in dimension 8 using computer algebra.  This
research was supported by the John Templeton Foundation. The opinions
expressed in this publication are those of the author and do not
necessarily reflect the views of the John Templeton Foundation.

\end{document}